\batchmode

\documentclass[conference]{worldcomp}

\usepackage[american]{babel}
\usepackage[T1]{fontenc}
\usepackage{times}
\usepackage{caption}

\usepackage{textcomp}
\usepackage{epsfig,graphicx}
\usepackage{xcolor}
\usepackage{amsfonts,amsmath,amssymb}
\usepackage{fixltx2e} 
\usepackage{booktabs}

\title{\bf Batch Spreadsheet for C Programmers}           

\author{
{\bfseries Richard Perry}\\
Villanova University \\
Department of Electrical and Computer Engineering \\
Villanova, PA, USA 19085 \\
richard.perry@villanova.edu
}

\begin{document}

\maketitle                        

\begin{abstract}
A computing environment is proposed, based on batch spreadsheet processing,
which produces a spreadsheet display from plain text input files of commands,
similar to the way documents are created using LaTeX. 
In this environment,
besides the usual spreadsheet rows and columns of cells,
variables can be defined and are stored in a separate symbol table.
Cell and symbol formulas may contain cycles, 
and cycles which converge can be used to implement iterative algorithms.
Formulas are specified using the syntax of the C programming language,
and all of C's numeric operators are supported, with operators such as ++,
+=, etc.\ being implicitly cyclic.
User--defined functions can be written in C and are accessed using a
dynamic link library.
The environment can be combined with a GUI front-end processor
to enable easier interaction and graphics including plotting.
\end{abstract}

\vspace{1em}
\noindent\textbf{Keywords:}
 {\small  Computational models, Software architectures for scientific computing}


\section{Introduction}

The idea for a batch spreadsheet processor was originally proposed in
\cite{fie2008} in the context of a course on parser design.
The evolution of that idea presented here, which goes far beyond
what was originally envisioned and implemented in the course,
is based on the premise of enabling anything that seems reasonable,
even if the usefulness is not immediately apparent.  So, for example,
if formula syntax is based on the C programming language,
then it should support all of C's operators, including ++, +=, etc.
This is discussed further in Section \ref{sec:cycles} on cycles.
Other examples, where the usefulness is apparent and the behavior
intuitively expected, are allowing lists (such as function arguments)
to be specified using cell ranges, cell ranges to be specified
using single cells (for a single--cell range), and ranges
to overlap and be specified in reverse and by columns or rows
as discussed in Section \ref{sec:ranges}.

After a brief review of the history of spreadsheets, we will present
the specifications for the proposed batch processor, together with
some examples from a partial implementation.

\section{Spreadsheet History}

The history of electronic spreadsheets \cite{ss25}
starts with Richard Mattessich \cite{RM} in 1964.
The output of his batch processor looked like a spreadsheet, 
but the input was purely data and all functionality was encoded
in the FORTRAN program implementing the business budget model.
In all subsequent spreadsheet implementations,
including the one proposed here,
the input specifies both the data and the functional
relationships among the cells.  

The modern interactive spreadsheet era starts in 1979 with
Visicalc, by Dan Bricklin and Bob Frankston. In 1982
James Gosling created the Unix \emph{SC} spreadsheet calculator
which has evolved through the work of others and is still in
use today.  In 1983 came Lotus 1-2-3, by Mitch Kapor, and
in 1985 Excel for Macintosh, by Microsoft (Windows came later).
There was also Quattro Pro, by Borland, in 1987, but in the battle for
the commercial spreadsheet market eventually Excel won.
Other popular opensource spreadsheet implementations include
Oleo (1992) and Gnumeric, from the GNU Project, 1998,
and OpenOffice, from StarOffice and Sun, 2000.

It is well-known that spreadsheet design and use can be
very error-prone \cite{corecalc,classsheets}, and most of the problems stem from
the interactive nature of modern spreadsheet features such as
cut/paste, adding and removing rows and columns, etc. 
The batch approach eliminates many of the traditional problems,
while enabling more powerful command-line types of options.
For example, to copy the entries from one range of cells to another,
a copy command like \emph{copy a2:c6 d6:b2} allows overlapping and
reverse ranges.  A batch spreadsheet processor can also be written using
portable C code, with no system-dependent GUI.
For ease of discussion, we will refer to the batch spreadsheet for C
programmers as simply ``SS'' for the rest of the paper.

\section{Input and Output}      \label{sec:IO}

SS first reads input from any files specified on the command line, then it
reads standard input for additional commands such as may be supplied by
a front-end GUI, or by the user directly.
The input may contain C or C++ style comments,
and macros using \#define.
As in C, input lines are joined together whenever one ends with a
backslash followed by a literal newline character.

Output goes to standard output by default, but can be redirected globally
using the output command, or redirected on a per--command basis using the
plot and print commands. ``stdout'' and ``-'', with or without quotes, can be
used as pseudo output file names to refer to standard output.

\section{Cell References}               \label{sec:cells}

Cell references may be specified in any one of three formats: A0, RC, and CR.
In A0 format, cells are specified by their column (one or two letters, A-Z,
case-insensitive) and row (one-to-three digits 0-9), with an optional \$
preceding the column and/or row value to indicate that the cell is fixed.
The column value A...ZZ represents an integer column number 0...701.
In RC and CR formats, cell are specified by their absolute row and column numbers,
using the letters R and C (or r, c), with brackets around relative
offsets.

When copying formulas, relative cell references remain relative to the
destination cells, and fixed references remain fixed.
The format for displaying cells in formulas may be selected using the format command.

\section{Ranges}                \label{sec:ranges}

A range consists of two cells, the start and end cells of the range, separated by a colon.
For example, A0:B9 (or A0:C1R9, or R0C0:R9C1, or C0R0:R9C1, etc.) specifies a range including rows 0
to 9 of columns A and B. The range start and end values do not have to be in increasing order;
B9:A0, B0:A9, and A9:B0 all refer to the same group of cells as A0:B9, but correspond to different
directions for traversing the range. For example the command
\emph{copy a0:a9 b9:b0} would copy column b to a in reverse order.

By default, ranges are traversed byrows to improve cache performance, since elements in a row are
adjacent in memory. That is, in pseudo-code:{\small
\begin{verbatim}
  for row = start_row to end_row
    for col = start_col to end_col
      use cell[row,col]
\end{verbatim}}

The bycols option can be used with the copy, eval, fill, plot, and print commands to cause evaluation by
columns. That is, in pseudo-code:{\small
\begin{verbatim}
  for col = start_col to end_col
    for row = start_row to end_row
      use cell[row,col]
\end{verbatim}}

Note that the starting row may be less than, equal to, or greater than the ending row. Same for
columns. So a range may consist of a single cell, row, or column (a0:a0, a0:d0, a0:a4), cells in
``forward'' order (a0:b4), or cells in ``partial reverse'' order (a4:b0, b0:a4), or cells in
``reverse'' order (b4:a0).

A range basically represents a list of cells, and is explicitly converted to a list when used in a
range assignment or numeric function argument.
A range consisting of a single cell may be specified using just that one cell, e.g.\ A0 as a range is
the same as A0:A0.

\section{Primitive Data Types}          \label{sec:data}

The SS primitive data types are double precision floating point, string, and constant.
All numeric calculations are performed and stored using double precision floating point.
A string is a sequence of characters enclosed in single or double quotes. No escape
sequences are recognized. If a string appears in a numeric calculation it is treated as
having the value 0.
The built-in constants are: HUGE\_VAL = Inf; RAND\_MAX = 32767;
pi = 3.14159 (computed from 4*atan(1)).
The values of the constants shown are from a Solaris/sparc system and may vary depending on
the system. Undefined cells are treated as having the value 0.

\section{Symbols}               \label{sec:symbols}

User-defined variables are stored in a symbol table and their formulas are evaluated each time the spreadsheet is
evaluated. Note that cell names can not be used as variables.
See Sections \ref{sec:example} and \ref{sec:cycles} for examples using symbols.

\section{Operators}             \label{sec:ops}

SS operators have the same precedence and associativity as those in the C
programming language. They mostly have the same meaning too, except for some of
the bitwise operators. The keywords NOT, AND, XOR, and OR, case--insensitive,
may also be used to represent the logical operators.  The bit shift operators
\verb%<<%, \verb%<<=%, \verb%>>%, and \verb%>>=% are implemented for
floating-point using ldexp() to adjust the binary exponent by the specified
power of 2.  Additionally, C's bitwise operators are treated as logical
operators, so \verb%~%, \&, \verb%^%, \verb%|% are the same as !, \&\&,
\verb%^^% (logical XOR), \verb%||% respectively, and \&=, \verb%^=%, \verb%|=%
are the same as \&\&=, \verb%^^=%, \verb%||=%.

\section{Numeric Functions}             \label{sec:nf}

SS includes all of the functions from the standard C math library, and more.
It includes the time function from time.h, rand and srand from stdlib.h,
and also irand, drand, and nrand functions for integer, uniform(0,1),
and normally distributed pseudo--random values respectively.

Most of the numeric functions take one expression argument and return one value.
A few of the functions take no arguments (rand, irand, drand, nrand, time),
and some take two arguments (atan2, fmod, frexp, ldexp, modf, pow).
For frexp and modf, the second argument must be a cell or symbol,
since it will be assigned one of the result values.

\section{Range Functions}               \label{sec:rf}

Range functions take an argument list of expressions and ranges and return one value
computed from the defined cells.
The range functions include:{\small
\begin{verbatim}
  avg       average
  count     number of cells defined
  majority  non-zero if majority are non-zero
  max       maximum
  min       minimum
  prod      product
  stdev     standard deviation
  sum       sum
\end{verbatim}}
User--defined numeric and range functions can be written in C
following a provided template to access SS internal data structures.
These are loaded at run--time from a dynamic link library.

\section{Commands}              \label{sec:commands}
Commands have their own syntax, consisting of keywords and options.
For all commands which traverse a range the \verb%byrows% or \verb%bycols%
option may be used to set the direction.
SS commands include:{\small
\begin{verbatim}
  byrows|bycols - set default direction
  copy dest_range src_range
  debug [on|off]
  eval [range|symbols] [number_of_iterations]
  exit
  fill range start_expr, increment_expr
  format A0|RC|CR - formula printing format
  format [cell|row|col|range] "fmt_string"
  load "fname"...
  output "fname"
  plot|plot2d|plot3d ["fname"] [range]
  print ["fname"] [range] [all|
    macros|symbols|formulas|values|
    formats|pointers|constants|functions]...
  quit
  srand expr - initialize rand()
\end{verbatim}}
If no range or symbols options are specified, eval evaluates the spreadsheet for the number of
iterations specified (default is 2 iterations). Each iteration first evaluates
the symbol table, then evaluates the cells twice: first starting at the top-left corner of
the cells being used and traversing the range to the bottom-right corner of the cells being
used; then again starting at the bottom-right corner and traversing to the top-left corner.
This catches most forward and reverse formula dependencies.
If the symbols option is specified, only the symbol table is evaluated for the number of
iterations specified.
If a range is specified, only the symbol table and that range are evaluated
for the number of iterations specified.

The fill command fills a range with constant values, starting with the start expression value, and increasing
by the increment expression value for subsequent cells. The start and increment expressions
are evaluated only once, before filling starts.

The format A0, RC, and CR options specify the format used for printing formulas.
For printing values, the format can be set globally or for a specific cell, row, column or
range. The default global format is ``\%.2f''. If a cell is not assigned a format, printing will
use the cell's row format, if set; otherwise it will use the cell's column format, if set;
otherwise it will use the global format.

The plot commands do not actually plot anything, they simply display output in a form
suitable as input to another program like a plot utility.

\section{Example}               \label{sec:example}
The following example
shows a set of student scores being scaled by their average and standard
deviation. Columns A and B represent the ``natural'' use of a spreadsheet for lists
of values related to or computed from a parallel list of values.  Cells C1 and D1 represent
an ``unnatural'' situation which can occur frequently in spreadsheets where we need
to store values somewhere, but their exact position does not matter relative to anything else.
It would be more natural to store these types of values in variables instead of a spreadsheet
cell.  The example does that for the computation of the mean, in variable \emph{mean},
and the use of that variable in the formula for column A is more ``natural'' and readable
than the use of \$d\$1 for the standard deviation.
{\small
\begin{verbatim}
% cat grades.ss
a0:d0 = { "grade", "score", "avg", "stdev"};
mean=avg(b1:b5); c1=mean; d1=stdev(b1:b5);
a1=80+15*(b1-mean)/$d$1; copy a2:a5 a1:a4;
b1:b5 = { 57, 67, 92, 87, 76 }; eval;
print symbols values formulas pointers;
% ss < grades.ss

mean = avg(B1:B5) = 75.8    (symbols)

        A       B       C       D
0       grade   score   avg     stdev
1       60.29   57.00   75.80   14.31
2       70.77   67.00
3       96.98   92.00       (values)
4       91.74   87.00
5       80.21   76.00

        A
0       "grade"
1       80+((15*(B1-mean))/$D$1)
2       80+((15*(B2-mean))/$D$1)
3       80+((15*(B3-mean))/$D$1)
4       80+((15*(B4-mean))/$D$1)
5       80+((15*(B5-mean))/$D$1)

        B       C       D
0       "score" "avg"   "stdev"
1       57      mean    stdev(B1:B5)
2       67
3       92                  (formulas)
4       87
5       76

        A       B       C       D
0       1058e28 1058e60 1058e98 1058ed0
1       1059288 0       1058fb0 1059058
2       1059288 0
3       1059288 0           (pointers)
4       1059288 0
5       1059288 0
\end{verbatim}}
From the ``print pointers'' output at the end
we can see that cells A1:A5 all refer to the same formula.

\begin{figure}[ht]
\begin{center}
\includegraphics[scale=0.4,angle=0]{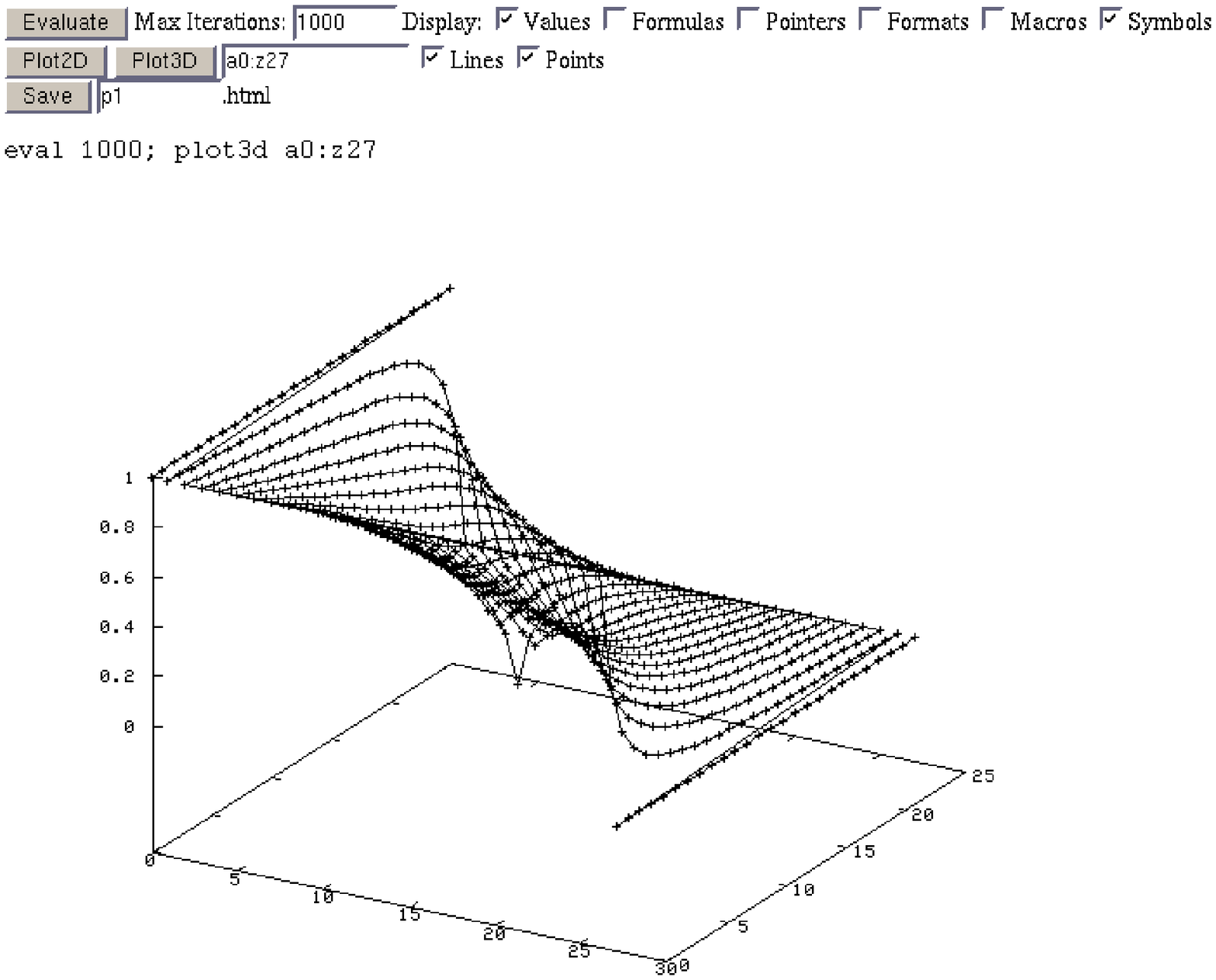}
\caption{Spreadsheet GUI with 3D Plot}
\label{fig:plot3d}
\end{center}
\end{figure}

\section{Cycles and Convergence}                \label{sec:cycles}

If a cell depends on itself, that forms a cycle and the spreadsheet may not converge when evaluated.
Cycles which converge can be used to implement iterative algorithms. For example, the following
spreadsheet uses Newton's method to find the square root of x:{\small
\begin{verbatim}
% cat sqrt.ss
x = 2; a0 = b0 ? b0 : x/2; b0 = (a0+x/a0)/2;
format "%20.18g";
\end{verbatim}}
Since a0 depends on b0, and b0 depends on a0, there is a cycle.
a0 will be set to b0 if b0 is non-zero, otherwise a0 will be set to x/2 to initialize the algorithm.
So a0 represents the previous value of b0, and b0 represents the next estimate of the square root.
Newton's method converges quickly:{\small
\begin{verbatim}
% echo "print all; eval a0:b0 10; \
print values;" | ss sqrt.ss

x = 2

  A                    B
0 B0 ? B0 : ((x/2))    (A0+(x/A0))/2

  A                    B
0 0                    0

ss_eval:: converged after 7 iterations

  A                    B
0 1.41421356237309492  1.41421356237309492
\end{verbatim}}
Finite element analysis is another application which requires iteration and can be set up in a
spreadsheet. In the following small example, the value of each non-boundary cell is computed as the
average of the cell's four nearest neighbors.{\small
\begin{verbatim}
% cat cycles.ss
// average of 4 nearest neighbors
R1C1=(R[]C[-1]+R[]C[+1]+R[-1]C[]+R[+1]C[])/4;
copy r1c2:r1c5 r1c1:r1c4;// set up one row
copy r2c1:r5c5 r1c1:r4c5;// copy to rows 2..5
fill r0c0:r0c6 1, 0;// boundary conditions,
fill r1c0:r6c0 1, 0;//  1's top and left
fill r1c6:r6c6 0, 0;//  0's right and bottom
fill r6c1:r6c5 0, 0;
format "%6.4f"; format RC; print values;
eval; eval 1000; print values;
% ss < cycles.ss
     0       1       ...     5       6
0    1.0000  1.0000  ...     1.0000  1.0000
1    1.0000  0.0000  ...     0.0000  0.0000
2    1.0000  0.0000  ...     0.0000  0.0000
3    1.0000  0.0000  ...     0.0000  0.0000
4    1.0000  0.0000  ...     0.0000  0.0000
5    1.0000  0.0000  ...     0.0000  0.0000
6    1.0000  0.0000  ...     0.0000  0.0000
ss_eval: still changing after 2 iterations
ss_eval: converged after 74 iterations

     0       1       ...     5       6
0    1.0000  1.0000  ...     1.0000  1.0000
1    1.0000  0.9374  ...     0.5000  0.0000
2    1.0000  0.8747  ...     0.2990  0.0000
3    1.0000  0.8040  ...     0.1960  0.0000
4    1.0000  0.7010  ...     0.1253  0.0000
5    1.0000  0.5000  ...     0.0626  0.0000
6    1.0000  0.0000  ...     0.0000  0.0000
\end{verbatim}}

A larger finite element analysis example is shown in Figure \ref{fig:plot3d}
using a web--based GUI front-end with 3D plotting.

\begin{figure}[ht!]
\includegraphics[scale=0.40,angle=0]{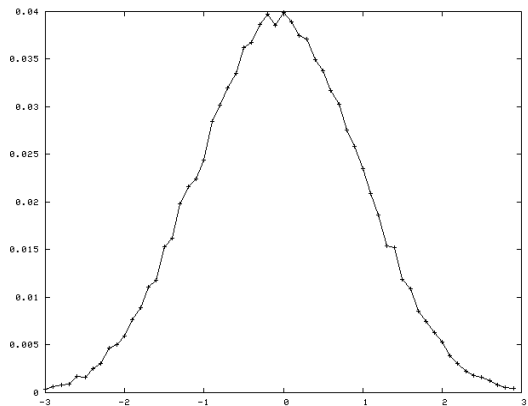}
\caption{Testing the Normal Distribution}
\label{fig:nrand}
\end{figure}

The spreadsheet may not converge when using operators ++, \verb%--%, +=, *=, etc.\ and the
pseudo-random number generator functions, since they produce varying values on each
evaluation. However, these operators and functions are still useful, in particular for Monte-Carlo
simulations.
The following simple example generates pseudo-random values for a0, with b0 representing the sum, c0
the evaluation count, and d0 the average:{\small
\begin{verbatim}
% cat rand.ss
a0 = drand(); b0 += a0; d0 = b0/++c0;
eval a0:d0 10; print values;
% ss < rand.ss
ss_eval:: still changing after 10 iterations
        A       B       C       D
0       0.37    4.90    10.00   0.49
\end{verbatim}}
The following example contains cycles in the formulas for symbols
\emph{sample} and \emph{trials}, and cells \emph{c1:c60}.
It tests the distribution of nrand values
using 60 counters over the range -3 to 3 with 50000 samples (iterations).{\small
\begin{verbatim}
sample = nrand(); trials += 1;
fill a1:a61 -3, 0.1;
c1 += (sample>=a1)&&(sample<a2) ? 1 : 0;
b1 = c1/trials; copy b2:c60 b1:c59;
eval c1:b60 50000; plot a1:b60;
\end{verbatim}}
The results, plotted in Figure \ref{fig:nrand}, are consistent with
the normal distribution.

\section{Omissions}	\label{sec:omit}

SS intentionally omits certain features which are commonly found in spreadsheets
and other computational environments.
It omits data types other than simple strings and double precision floating point,
targeting mainly numerical applications involving floating point operations.
Internally, other data types may be used as appropriate when implementing
specific algorithms, but the cell and symbol values are stored using floating point.

It also omits matrices and matrix operations, on the basis that applications
requiring matrix operations are best implemented using one of the many existing
high--quality commercial or free software packages designed for those kinds of applications.

\section{Extensions}	\label{sec:ext}

The current implementation of SS contains less than 40 numeric and range functions,
which represent the most basic framework necessary for proof--of--concept.
Most spreadsheets and programming environments contain hundreds of functions;
for example, Gnumeric \cite{gnumeric} contains 520, so any practical implementation
must allow for easy extension by adding new functions.

The implementation of new internal functions should follow the same templates
as user--defined functions,
\newpage
\noindent
the main difference being that internal functions
are compiled--in whereas user--define functions are dynamically linked in at run--time.
For example, the complete implementation of the irand function is:{\small
\begin{verbatim}
/* 1 pseudo-random integer, 0<=irand(i)<=i-1
*/
double nf_irand(const Node *n, const Cell *c)
{
  int i = eval_tree( n->u.t.right, c);

  return (int) (i*(rand()/(RAND_MAX+1.0)));
}
\end{verbatim}}
By simply placing that code in an irand.c file in the numeric function sub--directory
of the source code, it will be compiled--in.  The first--line comment is used
to specify how many arguments the function takes and a description that will be
included in both the run--time help command and the documentation.
The template for all functions consists of parse tree node and reference cell arguments,
and a double return value.

\section{Conclusion}

The technical literature on spreadsheet implementation
is relatively sparse \cite{corecalc}, as opposed to publications covering use of
spreadsheets.
Innovation is hampered by requiring Excel compatibility
and designing for non--programmer users.
The computing environment proposed here brings together the
spreadsheet and workspace (variables and symbol table)
paradigms, with powerful batch input commands and C--compatible formula syntax.
It provides flexibility in formula evaluation and range traversal,
and allows cycles, so it can be used to implement iterative as well as
traditional spreadsheet applications.


\end{document}